\documentclass[journal]{IEEEtran}
\ifCLASSINFOpdf

\else

\fi
\usepackage[utf8]{inputenc}
\usepackage[english]{babel}
\usepackage{makeidx}
\usepackage{graphicx}
\usepackage{amsmath}
\usepackage[inline]{enumitem}
\usepackage{multirow}
\usepackage{subcaption}
\usepackage{caption}
\usepackage{color}
\usepackage{booktabs}
\usepackage{siunitx}
\usepackage{cite}
\usepackage{algpseudocode}
\usepackage{algorithmicx}
\usepackage{array}
\usepackage{float}
\usepackage{picins}
\usepackage{fontawesome}
\usepackage{todonotes}
\usepackage[linesnumbered,ruled,vlined]{algorithm2e}
\usepackage{amsfonts}
\usepackage{threeparttable}
\usepackage{ragged2e}
\DeclareUnicodeCharacter{2061}{}
\usepackage{footmisc}
\usepackage{scrextend}
\usepackage{blindtext}
\usepackage[hidelinks]{hyperref}

\newcommand{\TT}[1]{{\color{black}{#1}}}
\newcommand{\NW}[1]{{\color{black}{#1}}}
\newcommand{\RV}[1]{{\color{black}{#1}}}
\newcommand{\RVV}[1]{{\color{black}{#1}}}
\newcommand{\RVVV}[1]{{\color{black}{#1}}}
\newcommand{\R}[1]{{\color{black}{#1}}}

\begin{document}
%
\title{Incremental Cross-Domain Adaptation for Robust Retinopathy Screening via Bayesian Deep Learning}
%
%
%

\author{Taimur Hassan\textsuperscript{*},~\IEEEmembership{Member,~IEEE,} 
        Bilal Hassan,
        Muhammad~Usman~Akram,~\IEEEmembership{Senior Member,~IEEE,}
        Shahrukh Hashmi, 
        Abdel Hakim Taguri,
        Naoufel~Werghi,~\IEEEmembership{Senior~Member,~IEEE}
\thanks{\noindent © 2021 IEEE.  Personal use of this material is permitted.  Permission from IEEE must be obtained for all other uses, in any current or future media, including reprinting/republishing this material for advertising or promotional purposes, creating new collective works, for resale or redistribution to servers or lists, or reuse of any copyrighted component of this work in other works.}
\thanks{\noindent This work is supported by a research fund from Khalifa University: Ref: CIRA-2019-047, and the Abu Dhabi Department of Education and Knowledge (ADEK), Ref: AARE19-156.}
\thanks{\noindent T. Hassan and N. Werghi are with the Center for Cyber-Physical Systems (C2PS), Department of Electrical Engineering and Computer Science, Khalifa University, Abu Dhabi, United Arab Emirates.}
\thanks{\noindent T. Hassan and M. U. Akram are with the Department of Computer and Software Engineering, National University of Sciences and Technology, Islamabad, Pakistan.}
\thanks{\noindent B. Hassan is with the School of Automation Science and Electrical Engineering, Beihang University, Beijing, China.}
\thanks{\noindent S. Hashmi is with the Department of Internal Medicine, Mayo Clinic, Minnesota, USA}
\thanks{\noindent A. H. Taguri is with the Abu Dhabi Healthcare Company (SEHA), Abu Dhabi, United Arab Emirates.}
\thanks{\noindent \textsuperscript{*} Corresponding author, Email: taimur.hassan@ku.ac.ae}}

\markboth{IEEE Transactions on Instrumentation and Measurement, October 2021}
{Hassan \MakeLowercase{\textit{et al.}}: Incremental Cross-Domain Adaptation for Robust Retinopathy Screening via Bayesian Deep Learning}
\maketitle
\begin{abstract}
\R{Retinopathy represents a group of retinal diseases that, if not treated timely, can cause severe visual impairments or even blindness. Many researchers have developed autonomous systems to recognize retinopathy via fundus and optical coherence tomography (OCT) imagery. However, most of these frameworks employ conventional transfer learning and fine-tuning approaches, requiring a decent amount of well-annotated training data to produce accurate diagnostic performance. This paper presents a novel incremental cross-domain adaptation instrument that allows any deep classification model to progressively learn abnormal retinal pathologies in OCT and fundus imagery via few-shot training. Furthermore, unlike its competitors, the proposed instrument is driven via a Bayesian multi-objective function that not only enforces the candidate classification network to retain its prior learned knowledge during incremental training but also ensures that the network understands the structural and semantic relationships between previously learned pathologies and newly added disease categories to effectively recognize them at the inference stage. The proposed framework, evaluated on six public datasets acquired with three different scanners to screen thirteen retinal pathologies, outperforms the state-of-the-art competitors by achieving an overall accuracy and F1 score of 0.9826 and 0.9846, respectively. }

\end{abstract}

\IEEEpeerreviewmaketitle

\section{Introduction}
\label{S:1}
\noindent \IEEEPARstart{T}{he} human eye consists of three layers, where the retina is the innermost layer responsible for producing vision. Retinal diseases or retinopathy tend to damage the retina resulting in a severe loss of vision or even blindness \cite{tim1}. Some of the serious retinal diseases are diabetic macular edema (DME), age-related macular degeneration (AMD), and central serous retinopathy (CSR). \RV{DME is caused by hyperglycemia (diabetes), where blood vessels become thinner and start leaking fluid deposits within the retina \cite{hassan2021BSPC}} According to the Early Treatment Diabetic Retinopathy Study (ETDRS), DME is graded as clinically significant macular edema (CSME) if 1) the retinal thickening is present within 500$\mu$m of the macular center, 2) hard exudates are discovered within 500$\mu$m of the macular center with adjacent macular thickening, and 3) there is a retinal thickening of one or more disc diameters and part of this thickening is within one disc diameter of the macular center \cite{etdrs}. Otherwise, it is graded as non-clinically significant macular edema (non-CSME). Similarly, the severity of CSME is further graded as centrally-involved DME (ci-DME) if retinal thickening is observed within the central sub-field zone of the macula (having diameter $\geq$ 1mm) in optical coherence tomography (OCT) scans. Otherwise, DME is classified as non-centrally involved (nci-DME) \cite{transferability}. 
\noindent AMD is another retinal condition (mostly found in elder people) that causes severe visual impairments if not treated timely. AMD is typically graded into two stages, i.e., the dry AMD and the wet AMD. Dry AMD is an early stage in which the retinal pigment epithelium (RPE) layer starts to degenerate, producing drusen and causing the subjects' central vision to become vivid and twirled. With the disease's progression, the abnormal blood vessels start to grow from the choroid and intercept \RV{the} retina, producing retinal fluids and other chorioretinal abnormalities such as scars and choroidal neovascularization (CNV). This stage is typically graded as wet AMD. 
CSR is another retinal syndrome that is \RV{mainly} caused due to stress. Clinically, CSR is diagnosed by observing the formation of serous detachment beneath the retina, and it is graded as acute, acute-persistent, and chronic. Acute CSR is an early stage where fluid-filled serous detachment forms beneath the retina (near to fovea). It becomes persistent if the serous detachment lasts for more than three months, which eventually leads to long-lasting chronic CSR in which RPE starts to degenerate and leads towards the formation of fibrosis, and CNV \cite{ragfw}.  

\noindent The retinal pathologies can be identified in a non-invasive manner through retinal fundus \cite{tim5} and OCT examinations \cite{tim33}. OCT imagery is more advantageous than fundus imagery as it gives a cross-sectional view of the retina for early disease identification. Also, OCT imagery can aid ophthalmologists in objectively assessing the severity of the underlying disease, resulting in a quick, accurate, and objective diagnosis. Nevertheless, the significance of fundus imagery for screening and grading certain retinal diseases cannot be fully ignored \cite{review}. Also, OCT imagery cannot identify blood and is limited towards detecting diseases that involve bleeding retina \cite{retinalImaging}. In such cases, fundus imagery provides an excellent alternative, aiding the doctors in analyzing the underlying pathologies. 

\NW{
\subsection{Motivation} \noindent Many researchers have developed novel solutions for extracting retinal layers \cite{cbm}, and retinal lesions \cite{Hassan2021TIM} for the lesion-aware screening \cite{Yoo2019MBEC} and grading of retinopathy \cite{LACNN}. Most of these frameworks utilize standard pre-trained networks in a transfer learning mode, which eliminates the need for training the model from scratch \cite{cbm}. However, the transfer learning and fine-tuning models have an inherent limitation of forgetting the prior knowledge upon learning new target-domain tasks \cite{hassan2021cbm}, and this limitation constrain them to identify only a limited number of pathologies. Scaling up these models to accommodate new disease patterns across different modalities requires computational, resource-demanding, and explicit re-training routines \cite{Yoo2019MBEC, Hassan2019Sensors}.
\R{Moreover, re-training these models again (to overcome scanner or pathological differences) is also an infeasible option for the clinicians at the hospitals and clinical setups. To cater these issues, we present a novel incremental cross-domain adaptation instrument that, with few-shot training, enables the deep classification models to recognize different retinal pathologies (across multi-modal imagery) irrespective of the scanner specifications or the pathological differences. Furthermore, the proposed framework can be easily tailored (in clinical settings) to recognize additional disease variants (while distilling its previously learned knowledge) due to its capacity to analyze the mutual contextual, structural, and spatial differences between incrementally learned knowledge representations via Bayesian inference.}
}

\section{Related Work} \label{sec:relatedWork}

\noindent Retinal image analysis is a widely researched topic \cite{review}, where researchers have developed various solutions to analyze retinal layers \cite{syed2016CMPB},  retinal lesions \cite{LACNN, Hassan2016JOSAA} via fundus \cite{VaradarajanNatureComm}, OCT  \cite{rabbani}, and fused fundus and OCT imagery \cite{transferability}. In this section, we first categorize the literature based upon their conventional deep learning approaches. Afterward, we shed light on some of \RV{the} recent frameworks which use advanced deep learning schemes to recognize retinal disease patterns. 

\subsection{Conventional Deep Learning Approaches}
\noindent 
Deep learning has been extensively utilized to detect normal and abnormal retinal pathologies \cite{Bilal2021CBM}. Kermany et al. \cite{zhang} pioneered these efforts by developing a CNN-based referral framework to predict AMD, CNV, DME, and normative pathologies from the macular OCT scans \cite{zhang}. Furthermore, they released their dataset publicly, which is one of the largest retinal OCT datasets to date. Rong et al. \cite{Rong2019JBHI} proposed a surrogate-assisted CNN classification model to recognize AMD, DME, and normal pathologies depicted within the retinal OCT scans. 
Lee et al. \cite{lee2017Ophthalmology} developed a deep learning system to automatically screen healthy and AMD pathologies. Arcadu et al. \cite{Arcadu2019Nature} developed a CNN coupled random forest to screen and grade diabetic retinopathy from fundus imagery.  Yoo et al. \cite{Yoo2019MBEC}  proposed a multi-modal approach whereby a pre-trained VGG-19 \cite{vgg} is used to extract feature representations from both fundus and OCT imagery to predict AMD pathologies. Apart from this, researchers have also devised hybrid CNN models for lesion-aware screening \cite{LACNN} and grading \cite{tbme} of retinopathy from both fundus \cite{Wei2020ICPR}, and OCT imagery \cite{ragfw}. 

\subsection{Advanced Deep Learning Methods}
\noindent To overcome the requirement of a large-scale and well-annotated training data, researchers have developed advanced deep learning approaches employing incremental learning \cite{Fu2021OphthalmologyRetina}, multi-task learning \cite{Wang2019MICCAI},  meta-learning \cite{Hasan2020Meta}, and domain adaptation \cite{d4}. 
Similarly, many researchers have turned their attention towards utilizing these schemes for extracting retinal lesions and screening retinal pathologies \cite{ADINet}. In this context, the work of He et al. \cite{He2020DA} is particularly appreciable as they utilized unsupervised domain adaptation, via adversarial learning, to segment retinal boundaries from the healthy OCT B-scans, which are acquired through Cirrus and Spectralis machines. 
Hasan et al. \cite{Hasan2020Meta} utilized model agnostic meta-learning \cite{maml} to register multi-vendor retinal images by refining a transformation matrix obtained through the spatial transformer in an adversarial manner \cite{Hasan2020Meta}. Similarly, Ju et al. \cite{Ju2020arXiv} utilized adversarial domain adaptation for achieving cross-domain shifts between normal and ultra-wide-field fundus photography to diagnose diabetic retinopathy, AMD, and glaucoma. Meng et al. \cite{ADINet} developed an attribute-driven incremental network (ADINet) to learn retinal diagnostic tasks from the fundus imagery incrementally.  

\noindent The above-mentioned approaches have tried to overcome the requirement of large-scale and well-annotated data for training deep learning models. However, to the best of our knowledge, there is no scheme till now that allows the classification networks to incrementally learn retinopathy screening tasks across multi-modal domains via few-shot training and perform these screening tasks simultaneously, at the inference stage, by analyzing the domain-specific abnormal retinal attributes. Such a scheme is highly desirable in clinical practice, as it not only saves time for re-training the model repeatedly on the large-scale data to learn new disease patterns, but it also allows the clinicians to train a single model, with few training examples, to recognize rarely occurring retinal diseases which are observed across the multi-modal imagery.  

\begin{figure*}
\begin{center}
\includegraphics[width=1\linewidth]{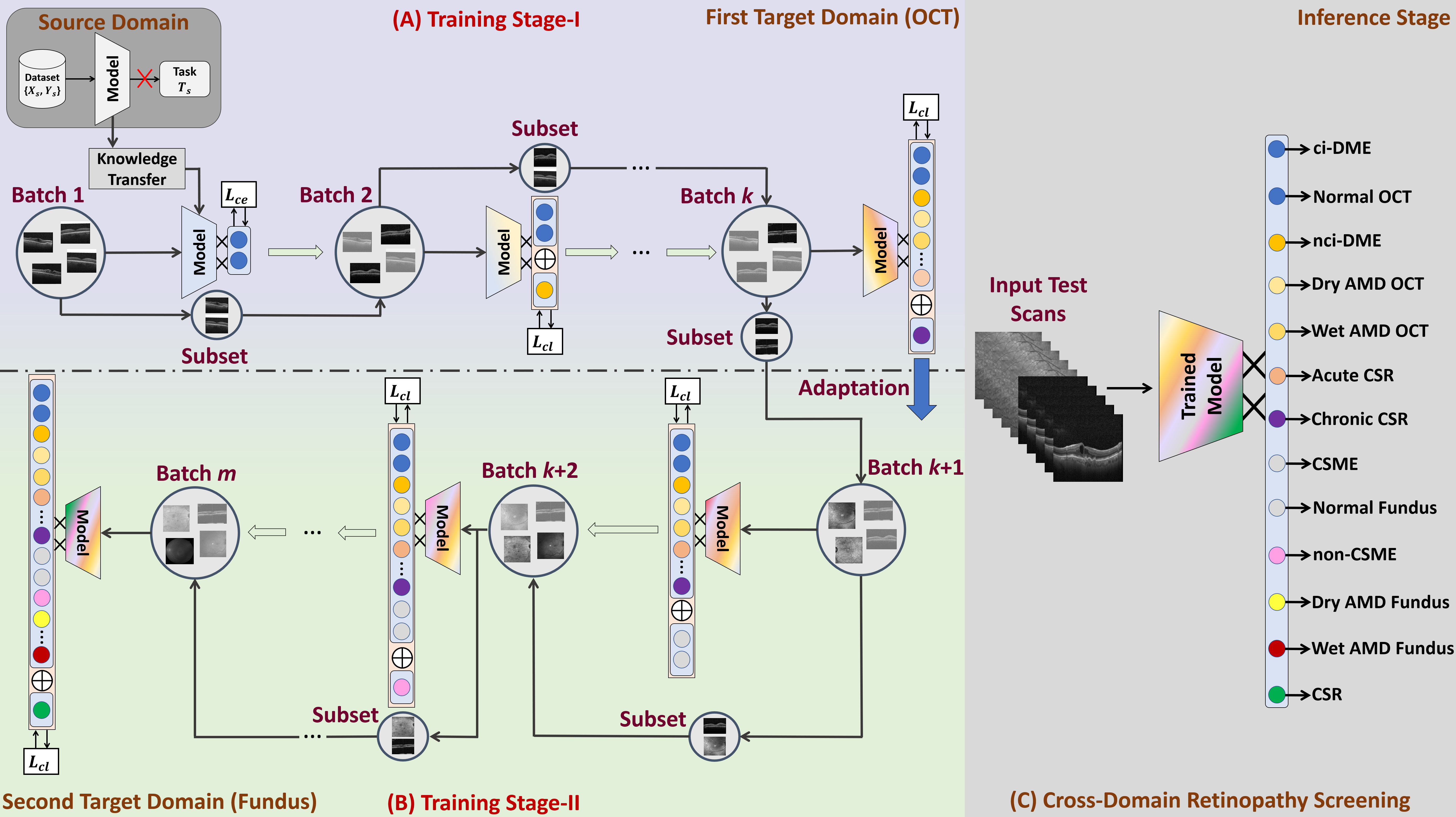}
\end{center}
   \caption{\R{Block diagram of the proposed framework. The first training stage is shown in (A), in which the classification network is incrementally trained on the first target domain to recognize $k+1$ retinal pathologies from the OCT scans. (B) denotes the second training stage in which we incrementally adapt the network to recognize $m+2$ retinal pathologies from the fundus imagery. During the inference stage (C), for $k=6, m=11$, the proposed incrementally trained cross-domain classifier can mass-screen thirteen retinal pathologies from both fundus and OCT imagery simultaneously, irrespective of the scanner specifications.} }
\label{fig:short2}
\end{figure*}

\subsection{Contributions} \label{sec:contribution}
\noindent This paper presents an original incremental cross-domain adaptation approach that constrains the classification model via Bayesian inference to learn retinopathy screening tasks across multi-modal imagery in an incremental fashion. The proposed scheme eliminates the need for re-training or fine-tuning the model for learning new disease categories every time. This distinctive feature of our framework is quite relevant for screening rarely occurring diseases (such as different variants of CSR \cite{csrvariants}) for which procuring large-scale training data is both infeasible and impractical \cite{ragfw}. Furthermore, the proposed framework, unlike its competitors \cite{Yoo2019MBEC, Vaghefi2020Hindawei, Hassan2019Sensors}, eliminates the need for deploying separate models to screen retina across multiple modalities. Thus, it also overcomes the training overheads of explicitly tuning multiple networks. In addition to this, the proposed framework presents a scalable option, in clinical practice, for autonomous retinal screening using OCT and fundus imagery. To summarize, the main features of this paper are:

\begin{itemize}[leftmargin=*]
    \item This paper presents a novel scheme that exploits deep neural networks' tendency to incrementally learn cross-domain diagnostic tasks (via few training examples) and perform them simultaneously (at the inference stage) by paying attention to the disease-specific lesions across each modality.
    
    \item \TT{Furthermore, to the best of our knowledge, the proposed framework is the first attempt towards utilizing incremental cross-domain adaptation for screening retinal diseases across OCT and fundus imagery.}
    
    \item The proposed incremental cross-domain adaptation is achieved by periodically minimizing the proposed continual learning loss function ($L_{cl}$), which enables the classification model to screen retinal diseases (from multiple modalities) by identifying their structural similarities and complex inter-dependencies via Bayesian inference.

\end{itemize}     

\noindent The rest of the paper is organized as follows: Section \ref{sec:proposed scheme} presents the details of the proposed framework, Section \ref{sec:expSetup} describes the experimental setup, Section \ref{sec:results} showcases the detailed evaluation results, Section \ref{sec:discussion} presents in-depth discussion about proposed framework and Section \ref{sec:conclusion} concludes the paper.

\section{Proposed Scheme} \label{sec:proposed scheme}
\noindent 
The block diagram of the proposed scheme is shown in Figure \ref{fig:short2}. \NW{Here, we can see that the candidate classification model recognizes the normal and abnormal retinal pathologies from both OCT and fundus imagery using a novel incremental cross-domain adaptation scheme based on Bayesian inference. The complete framework training is performed in two stages where, in the first stage, the classification model is incrementally trained with few OCT training examples to recognize ci-DME, normal OCT, nci-DME, dry AMD, wet AMD, acute CSR, and chronic CSR pathologies. Afterward, in the second stage, it is incrementally adapted to fundus imagery to recognize CSME, non-CSME, dry AMD, wet AMD, and CSR pathologies. \RV{We also wanted to highlight that our proposed scheme is different than the conventional incremental learning and domain adaptation (DA) approaches. In traditional incremental learning schemes, the model periodically learns new classification tasks (belonging to the same domain) without catastrophically forgetting its prior knowledge \R{\cite{ADINet}}. This is achieved by constraining the underlying model (in each training increment) via distillation-driven loss functions, which ensures that the model remembers its prior knowledge (through past episodic memories \cite{_16} or through previously fed examples \cite{_8}) while learning the newly added class representations.
Moreover, the weights of the incremental learning model are tuned in a way that they produce decent classification performance (for predicting both old and newly learned classes) at the inference stage. Apart from this, in conventional DA approaches, the model first learns the source domain tasks from the provided training examples and then adapts to the target domains to perform the same inter-related tasks by resolving the contextual and spatial differences between the source-target domain scans \R{\cite{fda, He2020DA}\footnote{\R{More details on transfer learning, incremental learning, and domain adaptation are presented in a supporting document within the source code repository.}}}. Contrary to this, in the proposed approach, the classification model is incrementally adapted to multiple target domains such that, with few-shot training, it can perform all the learned tasks simultaneously. The backbone of our proposed approach is a novel knowledge distillation strategy which, instead of treating the incrementally learned cross-domain class representations independent of each other, exploits their mutual relationships via Bayesian inference and make the model understand these relationships to maximize the classification performance (even when the model is trained with few examples).} Also, unlike the conventional DA schemes, the proposed scheme allows the underlying network to new cross-domain tasks which are not necessarily related to each other. For example, the classification model (trained using the proposed scheme) can easily screen the ci-DME and nci-DME pathologies from OCT scans while recognizing the non-CSME pathologies from the fundus scans (where ci-DME, nci-DME in OCT scans has no association with the non-CSME pathology as it can only be visualized in the fundus scans). Therefore, we dubbed the proposed scheme as incremental cross-domain adaptation. The detailed description of the proposed framework's training and testing are presented in the subsequent sections:    
}
\subsection{Training Stage-I}
\noindent In the first training stage, we pick the pre-trained network (the model previously trained on a source domain dataset, i.e., ImageNet \cite{imagenet}). Then, in each training increment, we update its final layers to recognize retinal pathologies from the OCT imagery, which is the first target domain. In the first iteration, we make the network learn normal and ci-DME pathologies using their respective training examples. Similarly, in the second iteration, we stack a new class,  representing 'nci-DME' pathology, in the final layer and constrain the model to recognize the nci-DME affected OCT scans.  The model in the second iteration effectively learns the nci-DME disease patterns while retaining its previous knowledge about normal and ci-DME pathologies. The same process is repeated till $k$ iterations, where the network is periodically trained to screen $k+1$ retinal pathologies. 
It should be noted here that in each iteration $j$ (where $j > 1$), the network is constrained through the proposed $L_{cl}$ loss function  (see Section \ref{sec:lossfunctions}) to identify new types of retinal pathology while retaining its prior knowledge. Moreover, instead of only using training examples for the newly added categories, we also feed the network with the subset of examples used in the $1^{st}$ to $(j-1)^{th}$ iteration. This enables the $L_{cl}$ function to distill the network's previous knowledge while constraining it to learn newly added disease categories. Furthermore, $L_{cl}$ (via $L_{md}$) ensures that the network learns the structural and semantic relationships between previously learned and newly stacked classes to recognize them correctly at the inference stage. More details about $L_{md}$ and $L_{cl}$ are presented in Sections (\ref{sec:lossfunctions}, \ref{sec:lmd}, and \ref{sec:lcl}). Moreover, at $j=1$, we only train the model using the categorical cross-entropy loss function ($L_{ce}$) since we don't have any old classes in the first iteration (from the source-domain), which we want the network to retain.

\subsection{Training Stage-II}
\noindent In the second stage, we adapt the model, trained on the OCT domain, to recognize the abnormal retinal pathologies from the fundus imagery. The adaptation is performed incrementally where the network in iteration $k+1$ is constrained to identify healthy and CSME pathologies from the fundus scans. The same process is repeated till the $m^{th}$ iteration, where the network is trained to screen up to $m+2$ retinal pathologies. Note that the model, in the second training stage, not only learns to identify retinal pathologies across two different types of imagery but it also learns to recognize similar disease patterns across both domains to analyze the underlying disease severity. For example, at the inference stage, for the DME patient, the proposed framework possesses the capacity to screen it as CSME and non-CSME symptomatic using the corresponding fundus scan. If it is CSME, then the proposed framework can further check whether the screened CSME case is ci-DME or nci-DME by searching for the presence of retinal lesions such as intra-retinal fluid (IRF) sub-retinal fluid (SRF) and hard exudates (HE) within the respective OCT volume. 
The model achieves such extended screening without requiring any additional data or re-training, while to achieve a similar behavior,  state-of-the-art works, based on conventional transfer learning, would have to be explicitly trained on the large well-annotated data \cite{Yoo2019MBEC, Hassan2019Sensors}. 

\subsection{Inference Stage}
\noindent After training the model till the $m^{th}$ incremental iteration, we use it to screen $m+2$ retinal pathologies at the inference stage. We highlight here that the proposed framework employs a single classification network to screen retinal diseases from OCT and fundus imagery by analyzing their modality-specific abnormal lesions. For example, if an OCT scan of a CSR patient is passed into the proposed system, it will be screened as acute CSR or chronic CSR by observing the presence of IRF, SRF, and CNV. Similarly, if a fundus scan of the same subject is passed into the system, then it will be screened as CSR by analyzing the appearance of fluid blisters.  Now, suppose the model has to be tuned to recognize more types of pathologies. In that case, we only need to run one more incremental training iteration in which we will first stack the categories of these new pathologies within the final classification layer of the network. Then, we can make the model learn these newly added disease categories while retaining its previous knowledge, using the proposed $L_{cl}$ loss function and a small set of training examples. Thus, the proposed scheme allows low-cost scalability for screening retinopathy in clinical settings.

\subsection{Loss Functions}\label{sec:lossfunctions}
\noindent During incremental training, we constrain the classification network to accurately learn the newly added classes while retaining its prior learned knowledge via the proposed continual learning objective function ($L_{cl}$). This function encompasses three sub-objective functions, namely $L_{o}$, $L_{n}$, and $L_{md}$.  $L_{o}$ allows the classification network to retain the prior knowledge by minimizing the prediction errors on the training examples of the old classes (learned in the previous $1$ to $j-1$ iterations). $L_{n}$ enables the classification model to learn new classes (added in the $j^{th}$ iteration) by minimizing the network loss on the training examples of the new classes. Both $L_o$ and $L_n$ update the network weights such that it learns the maximum amount of information while showing high resistance against catastrophic forgetting. The loss function $L_{cl}$ also analyzes the mutual relationship and structural inter-dependencies between the previously added and the newly stacked classes via an auxiliary $L_{md}$ sub-objective function-driven via Bayesian inference. 
$L_{md}$ gives more exposure to the underlying classification model to understand and learn the diverse ranging yet semantically related categories by exploiting their inter-dependencies.  To the best of our knowledge, all the state-of-the-art knowledge distillation schemes ignore this aspect, assuming the older and newer knowledge representations are independent of each other. 
More details about  $L_{md}$, $L_{o}$  and $L_{n}$ are exposed in the next sections.

\subsection{Mutual Distillation Loss Function} \label{sec:lmd}
\noindent Knowledge representations learned by the classification model in each iteration, during incremental training, are generally non-mutually exclusive \cite{Gal2017ICML}. \NW{For example: consider a fruit classification model that learns to classify \textit{oranges} in one iteration and then \textit{lemons} in the next iteration. Since both of the fruits are related to each other, therefore, the classifier should be constrained during the incremental training so that it can accurately recognize these fruits together. More specifically, it should not lose its previously attained knowledge of recognizing \textit{oranges} while learning about \textit{lemons}.  To address this, we propose $L_{md}$ loss function that exploits the mutual structural and semantic inter-dependencies between different knowledge representations (during the incremental training process) to enhance the classification performance of the candidate model, while simultaneously making it more resistant to the catastrophic forgetting phenomena.} In each iteration $j$, where $j > 1$, a deep network is fed with the training examples $x$ such that $x=[x_o, x_n]$ where $x_o$ denotes the training examples of the old classes (learned in $1$ to $j-1$ iterations) and $x_n$ denotes the training examples of the new classes (added in the $j^{th}$ iteration). The network produces the output logits $l(x)$ such that $l(x)=w\times f(x)+b$, where $w$ denotes the network weights, $b$ represents the biasing factor and $f(x)$ is the feature vector. The logits $l(x)$ are formed by $l(x)=[l(x_o), l(x_n)]$, where $l(x_o)$ and $l(x_n)$  represent the logits produced through the $x_o$ and $x_n$, respectively. Typically, in incremental learning systems, we divide these logits by a constant $\tau$ (i.e., $l^\tau(x) = \frac{l(x)}{\tau}$), so that they produce soft-target probabilities when passed through the activation function \cite{survey}. Moreover, the joint probability distribution between the scaled logits $l^\tau(x_o)$ and $l^\tau(x_n)$ given the outcome $c_i$ is obtained through Bayes Rule as expressed below:
\begin{equation}
    p(c_i| l^\tau(x_o),l^\tau(x_n)) =  \frac{p(l^\tau(x_o),l^\tau(x_n) | c_i) p(c_i)}{\sum_{c_k=0}^{c_o-1} p(l^\tau(x_o),l^\tau(x_n)|c_k) p(c_k)},
    \label{eq:label14}
\end{equation}
where $p(c_i)$ denotes the prior probability, $p(l^\tau(x_o),l^\tau(x_n) | c_i)$ represents the likelihood for the given class $c_i$ and $c_o$ represents the total number of previously learned classes. It should be noted that for a given class $c_i$, the joint distributions between $l(x_o)$ and $l(x_n)$ are symmetrical i.e. $p(c_i| l^\tau(x_o),l^\tau(x_n)) = p(c_i| l^\tau(x_n),l^\tau(x_o))$. From the above, the mutual distillation loss is computed through: 
\begin{equation}
\begin{split}
L_{md}=-\sum\limits_{h=0}^{c_o-1} t(x_{o,h}) \log(p(c_h  | l^{\tau}(x_o),l^{\tau}(x_n)) ,
\end{split}
\label{eq:label8}
\end{equation}

\noindent where $t(x_{o,h})$ represents the ground truth label (in one-hot notation) of the training sample ($x_o$) belonging to the previously learned $h^{th}$ class. 

\subsection{Continual Learning Loss Function} \label{sec:lcl}
\noindent The $L_{md}$ bridges the gap between old and newly learned classes by constraining the network to learn their mutual relations and dependencies. But to make the network aware of their exclusive characteristics, we jointly optimize $L_{o}$ and $L_{n}$ objective functions along with $L_{md}$. 
The $L_{o}$ and $L_{n}$ are expressed as:
\begin{equation}
L_{o} = -\sum\limits_{s=0}^{c_{o}-1} t(x_{o,s}) \log(p(l^\tau(x_{o,s}))),
\label{eq:label10}
\end{equation}

\noindent and 

\begin{equation}
\begin{split}
L_{n} = \sum\limits_{s=0}^{c_{n}-1} q(t(x_{n,s})) \log \left(\frac{q(t(x_{n,s}))}{p(l(x_{n,s}))}\right) .
\end{split}
\label{eq:label11}
\end{equation}
 We can observe here that $L_n$ gives more emphasis on learning each of the newly added category, in the current iteration, by computing the relative entropy between true and predicted probability distributions. However, $L_o$ computes the total entropy between the true and predicted distributions ensuring that the network does not forget the prior learned categories while learning newly added class representations. Moreover, $t(x_{o,s})$ and $t(x_{n,s})$ are the true $s^{th}$ class indicators for the training examples of the older and newly added classes ($x_o$ and $x_n$), respectively, $q(t(x_{n,s}))$ and $p(l(x_{n,s}))$ represents the actual and predicted distributions generated from $t(x_{n,s})$ and $l(x_{n,s})$ for the $s^{th}$ class, respectively. Apart from this, $p(l^\tau(x_{o,s}))$ represents the distribution of scaled logits $l^\tau(x_{o,s})$ (for the $s^{th}$ class) generated through the training samples of the old classes $x_o$. 
 
\noindent The continual learning loss functions  $L_{cl}$ is, therefore, defined as a linear combination of $L_{md}$, $L_{o}$  and $L_{n}$ as expressed below:
\begin{equation}
L_{cl} = \alpha L_{n} + \beta L_{md} + \gamma L_{o}     ,
\end{equation}
\noindent where $\alpha$, $\beta$ and $\gamma$ denote the loss weights. These weights are empirically chosen to be 0.25, 0.45, and 0.30, respectively. \RVVV{Due to space constraints, we refer the reader to the supplementary material to see the in-depth ablative experimentation for determining these hyperparameters.}
 
\section{Experimental Setup} \label{sec:expSetup}
\noindent \RV{This section reports} the detailed description of the datasets, the training and implementation protocols, and the evaluation metrics, which we used to evaluate the proposed framework.

\subsection{Datasets}

\noindent  We have used six public datasets containing retinal fundus and OCT imagery. These datasets includes the Rabbani \cite{rabbani} dataset, BIOMISA \cite{biomisa} dataset, Zhang dataset \cite{zhang}, Duke-I dataset \cite{duke1}, Duke-II dataset \cite{duke2}, and the Duke-III dataset \cite{duke3}. The detailed description of these datasets are given below:

\subsubsection{Rabbani}
Rabbani dataset is one of the unique retinal image databases \RV{that contains a total of 4,241 OCT and 148 fundus scans showing normal, ci-DME, nci-DME, and dry AMD pathologies. Moreover, the dataset is acquired using Spectralis, Heidelberg Inc. at Noor Eye Hospital, Tehran, Iran.} 

\subsubsection{BIOMISA}
BIOMISA dataset is another public dataset introduced by the Biomedical Image and Signal Analysis (BIOMISA) Lab at the National University of Sciences and Technology (NUST), Islamabad, Pakistan. \RV{BIOMISA dataset contains a total of 5,324 OCT and 115 fundus scans from 99 subjects either healthy or suffering from ci-DME, nci-DME, dry AMD, wet AMD, acute and chronic CSR. Apart from this, the scans within the BIOMISA dataset are acquired through Topcon 3D OCT 2000 in the Armed Forces Institute of Ophthalmology, Rawalpindi, Pakistan.}

\subsubsection{Zhang}
\RV{Zhang dataset is one of the largest retinal OCT datasets containing 109,309 scans which shows wet AMD (choroidal neovascularization), dry AMD (drusen), ci-DME, nci-DME, and healthy pathologies.} The dataset has been arranged in a way that 108,309 scans are to be used for training and 1,000 scans are to be used for testing purposes. Moreover, the scans in the Zhang dataset are acquired through Spectralis, Heidelberg Inc. Zhang dataset \cite{zhang} also contains CXRs to screen pneumonia subjects. 

\subsubsection{Duke-I}
The Duke-I dataset is one of the largest OCT datasets from the Vision \& Image Processing (VIP) lab, Duke University, USA, \RV{containing 38,400 scans from healthy (controlled) and dry AMD subjects. The scans are acquired through the Bioptigen OCT machine.} The dataset also contains ground truth annotations to verify the performance of the automated systems towards extracting the retinal layers.

\subsubsection{Duke-II}
\RV{The Duke-II dataset contains 610 OCT scans showing nci-DME and ci-DME pathologies.} The dataset was introduced in \cite{duke2} by VIP lab, and it contains detailed annotations for the retinal layers and fluids marked by two expert clinicians. Moreover, the scans within Duke-II are acquired through Spectralis, Heidelberg Inc. Also, in some experiments, we combined Duke-I and Duke-II datasets to evaluate further the proposed framework towards screening the retinal pathologies covered by both of these datasets. 

\subsubsection{Duke-III}
Duke-III is the third dataset from the VIP lab that we used in this research. \RV{The dataset contains 3,231 retinal OCT scans reflecting normal, ci-DME, nci-DME, and dry AMD pathologies.} The dataset is primarily designed for the classification tasks, and the scans within the dataset are organized with respect to the pathologies that they exhibit. Moreover, Duke-III dataset scans are also acquired through Spectralis, Heidelberg Inc.

\noindent Apart from this, we also got all the OCT and fundus scans (contained within these datasets) extensively annotated by the expert clinicians, where all of these markings have been released publicly for the research community within the source code repository\footnote{\R{\label{note1}Source codes and dataset annotations are available at \url{https://github.com/taimurhassan/continual_learning/}}}. \RVV{Also, it should be noted that the scans within each dataset were of different sizes. But in order to make these sizes consistent with the input layer of the classification networks, we have resized the scans (within each dataset) to a standard resolution of $224\times 224\times 3$. }

\subsection{Training and Implementation Details} \label{sec:trainingDetails}
\noindent The major benefit of the proposed scheme compared to the conventional fine-tuning approaches is its ability to utilize few training examples to learn the diversified retinopathy classification tasks. 
\noindent Moreover, the complete training of the proposed framework is performed in two stages. The first stage relates to typical class-incremental learning \cite{survey}, where the model is trained in each increment to distinguish between different types of retinal pathologies from the provided OCT examples. Unlike conventional fine-tuning systems, the proposed framework does not learn all the classes (retinal diseases) at once from the large well-annotated training data. Rather it learns them incrementally (across each dataset) by utilizing a significantly lesser amount of training examples. Incremental training in the first stage lasts for $k=6$ iterations, where the classification model periodically learns to screen $k+1$ retinal pathologies. Moreover, in the second stage, the model is incrementally adapted to fundus imagery to recognize $m-k+1$ retinal diseases, where $m=11$. 
\noindent The detailed description of training protocols for Stage-I and Stage-II are presented next.

\begin{table}[t]
    \centering
    \caption{\RV{Retinal scans that are used for training the proposed framework in each incremental iteration. The abbreviations are ST: Stage, IT: Incremental Iteration, and PTS: Previous Training Subset. PTS contains the small portion of scans that were used in the previous iterations to aid the network in distilling its prior learned knowledge.}}
    \begin{tabular}{ccccccccc}
        \toprule
        \RV{ST} & \RV{IT} & \RV{Dataset} & \RV{Training Scans}\\\hline
        \RV{I} & \RV{1} & \RV{Rabbani \cite{rabbani}} & \RV{300 (100 ci-DME \& 200 normal)} \\
        && \RV{BIOMISA \cite{biomisa}} & \RV{500 (200 ci-DME \& 300 normal)} \\
        && \RV{Duke-II \cite{duke2}} & \RV{100 (all ci-DME)} \\
        && \RV{Duke-III \cite{duke3}} & \RV{350 (100 ci-DME \& 250 normal)} \\
        && \RV{Zhang \cite{zhang}} & \RV{4000 (2k ci-DME \& 2k normal)} \\\cline{2-4}
        
        & \RV{2} & \RV{Rabbani \cite{rabbani}} & \RV{150 (all nci-DME)} \\
        && \RV{BIOMISA \cite{biomisa}} & \RV{250 (all nci-DME)} \\
        && \RV{Duke-II \cite{duke2}} & \RV{100 (all nci-DME)} \\
        && \RV{Duke-III \cite{duke3}} & \RV{150 (all nci-DME)} \\
        && \RV{Zhang \cite{zhang}} & \RV{2000 (all nci-DME)} \\
        && \RV{PTS} & \RV{525 (265 ci-DME \& 260 normal)} \\\cline{2-4}
        
        & \RV{3} & \RV{Rabbani \cite{rabbani}} & \RV{120 (all dry AMD)} \\
        && \RV{BIOMISA \cite{biomisa}} & \RV{120 (all dry AMD)} \\
        && \RV{Duke-I \cite{duke1}} & \RV{150 (all dry AMD)} \\
        && \RV{Duke-III \cite{duke3}} & \RV{100 (all dry AMD)} \\
        && \RV{Zhang \cite{zhang}} & \RV{1000 (all dry AMD)} \\
        && \RV{PTS} & \RV{790 (525 old \& 265 nci-DME)} \\\cline{2-4}
        
        & \RV{4} & \RV{BIOMISA \cite{biomisa}} & \RV{200 (all wet AMD)} \\
        && \RV{Zhang \cite{zhang}} & \RV{2500 (all wet AMD)} \\
        && \RV{PTS} & \RV{1062 (790 old \& 272 dry AMD)} \\\cline{2-4}
        
        & \RV{5} & \RV{BIOMISA \cite{biomisa}} & \RV{120 (all acute CSR)} \\
        && \RV{PTS} & \RV{1152 (1062 old \& 90 wet AMD)} \\\cline{2-4}
        
        & \RV{6} & \RV{BIOMISA \cite{biomisa}} & \RV{150 (all chronic CSR)} \\
        && \RV{PTS} & \RV{1252 (1152 old \& 100 acute CSR)} \\\hline
        
        \RV{II} & \RV{1} & \RV{Rabbani \cite{rabbani}} & \RV{40 (20 CSME \& 20 normal)} \\
        && \RV{BIOMISA \cite{biomisa}} & \RV{20 (10 CSME \& 10 normal)} \\
        && \RV{PTS} & \RV{1292 (1252 old, 20 CSME \& 20 normal)} \\\cline{2-4}
        
        & \RV{2} & \RV{Rabbani \cite{rabbani}} & \RV{10 (all non-CSME)} \\
        && \RV{BIOMISA \cite{biomisa}} & \RV{10 (all non-CSME)} \\
        && \RV{PTS} & \RV{1302 (1292 old, 5 CSME \& 5 normal)} \\\cline{2-4}
        
        & \RV{3} & \RV{Rabbani \cite{rabbani}} & \RV{7 (all dry AMD)} \\
        && \RV{BIOMISA \cite{biomisa}} & \RV{20 (all dry AMD)} \\
        && \RV{PTS} & \RV{1322 (1302 old, 20 non-CSME)} \\\cline{2-4}
        
        & \RV{4} & \RV{BIOMISA \cite{biomisa}} & \RV{5 (all wet AMD)} \\
        && \RV{PTS} & \RV{1337 (1302 old, 10 dry AMD)} \\\cline{2-4}
        
        & \RV{5} & \RV{BIOMISA \cite{biomisa}} & \RV{10 (all CSR)} \\
        && \RV{PTS} & \RV{1342 (1337 old, 5 wet AMD)} \\
        
        \bottomrule
    \end{tabular}
    \label{tab:data}
\end{table}

\begin{table}[t]
    \centering
    \caption{Training and testing scans utilized for evaluating the proposed framework on each individual dataset. \RV{The column 'Training' represents the number of scans which we used for few-shot training, \RVV{whereas the column 'AT' represents the actual number of scans that are to be used for training purposes as per the dataset standard.} }}
    \begin{tabular}{cccccc}
        \toprule
        Dataset & Scans & Scanner & \RV{Training} & \RV{Testing} & \RVV{AT} \\\hline
        Rabbani  & OCT & Spectralis & \RV{570} & \RV{3671} & \RVV{1061}\\
        & Fundus & & \RV{57} & \RV{91} & \RVV{37} \\\hline
        BIOMISA & OCT & Topcon & \RV{1340} & \RV{3984} & \RVV{3840} \\
        & Fundus & OCT 2000 & \RV{65} & \RV{50} & \RVV{90}\\\hline
        Duke-I  & OCT & Bioptigen & \RV{150} & \RV{38250} & \RVV{300} \\\hline

        Duke-II  & OCT & Spectralis  & \RV{200} & \RV{410} & \RVV{305} \\\hline

        Duke-III  & OCT & Spectralis & \RV{600} & \RV{2631} & \RVV{3048} \\\hline

        Zhang & OCT & Spectralis & \RV{9500} & \RV{99809} & \RVV{108309} \\

        \bottomrule
    \end{tabular}
    \label{tab:data2}
\end{table}

\subsubsection{Stage-I Training Details}
In the first stage, we pick the pre-trained model, update its last layer to recognize ci-DME and healthy pathologies (in the first iteration). Here, we feed the network with 5,250 training examples where 300 are taken from the Rabbani dataset, 500 are taken from \RV{the} BIOMISA dataset, 100 examples are taken from Duke-II, 350 examples are taken from the Duke-III dataset, and 4,000 examples are taken from Zhang dataset. After learning these pathologies, we stack \RV{a} new 'nci-DME' class for which we feed 150 examples from Rabbani, 250 examples from BIOMISA, 100 examples from Duke-II, 150 examples from Duke-III, and 2,000 examples for the Zhang dataset. Furthermore, we also feed the network with a small subset of 550 examples from the previous training batch to retain the previously learned classes. In the third iteration, we make the network learn the dry AMD-affected pathologies for which we feed 1,470 training examples from Rabbani, BIOMISA, Duke-I, Duke-III, and Zhang dataset. In the fourth iteration, we trained the network to identify wet AMD using 2,720 training examples from Rabbani, BIOMISA, and Zhang dataset. In the fifth and sixth iteration, we trained the classification model to identify acute and chronic CSR from \RV{the} BIOMISA dataset using 120 and 150 training examples, respectively. The reason for using only the BIOMISA dataset for identifying CSR is because this is the only public dataset that contains annotated CSR pathologies \cite{biomisa}.

\subsubsection{Stage-II Training Details}
After learning normal and abnormal retinal pathologies from the OCT scans, we adapt the classification model to learn about the retinal pathological variations from the fundus scans incrementally. Here, in the first iteration, we stack two classes representing healthy and CSME affected fundus, and the network learns these categories from a total of 40 training fundus scans from the Rabbani and BIOMISA datasets. Moreover, in the next incremental iterations, the proposed framework learns to screen non-CSME, dry AMD, wet AMD, and CSR pathologies using 20, 27, 5, and 10 training examples, respectively, as shown in Table \ref{tab:data}. 

\subsubsection{Incremental Training on Individual Datasets}
In order to compare the performance of the proposed framework with state-of-the-art retinal diagnostic systems, we also evaluated the proposed system on each dataset separately. The number of training iterations varies on each dataset depending upon the number of retinal pathologies and the modalities it contains. However, the number of training scans for each dataset is presented in Table \ref{tab:data2}. Here, we highlight that for each dataset, the proposed framework only utilizes a small subset of training data (which is defined by the dataset standard) to learn the retinal pathologies, and it also produces competitive performance against the state-of-the-art methods, which are trained in a conventional manner using the complete training data (as defined by each dataset protocol). 

\subsubsection{Implementation Protocols}
\noindent The proposed scheme is implemented using TensorFlow 2.1.0 with Keras 2.3.0 on the Anaconda platform with Python 3.7.9. Some of the utility functions have also been written in MATLAB R2020a.

\noindent Apart from this, each incremental training iteration lasts for five epochs (where the number of cycles in each epoch varies for each dataset). Moreover, the optimizer used during the training was ADADELTA \cite{Zeiler2012ADADELTA}, and the training was conducted on a machine with Core i7-9750H@2.6 GHz, 32 GB DDR4 RAM, and NVIDIA RTX 2080 Max-Q GPU. The source code of the proposed framework and its complete documentation is released publicly on GitHub\textsuperscript{\ref{note1}}.

\subsection{Evaluation Metrics} \label{sec:metrics}
\noindent Standard classification metrics, such as accuracy (Acc), true positive rate (TPR), true negative rate (TNR), positive predicted value (PPV), and F1 score, have been utilized to measure the performance of the proposed framework. Moreover, we also used the ROC curve and AUC to assess the behavior of the proposed framework towards screening the retinal pathologies acquired with specific types of scanners. \RVV{Apart from this, to measure the performance difference between the proposed framework and the state-of-the-art works, we used the relative percentage formula, i.e., $\frac{l_e-l_g}{l_g}\times 100$, where $l_e$ represents the leading score, and $l_g$ denotes the lagging score.}

\section{Results} \label{sec:results}   
\noindent In this section, we present a detailed evaluation of the proposed framework for screening retinopathy using both fundus and OCT imagery. In the first set of experiments, we compared the proposed framework's performance with the state-of-the-art retinal diagnostic systems on different publicly available datasets. Afterward, in the second set of experiments, we compared the proposed framework's performance with state-of-the-art domain adaptation and incremental domain adaptation schemes for performing the cross-domain retinopathy screening tasks. Before discussing these experiments, we present comprehensive ablation studies to determine the proposed system's hyperparameters and the optimal classification network. \RV{Apart from this, we give a detailed description of the number of scans which we used for training and testing purposes in Table \ref{tab:data} and \ref{tab:data2}.}

\subsection{Ablation Studies}
\noindent The goal of the ablative analysis is to: 1)  analyze the effect of temperature in each iteration for computing the soft probabilities and 2) to determine the optimal classification network. \RVVV{We also report additional ablation experiments in the supplementary material through which \RV{we determined the number of learnable parameters of each network} and the hyperparameters of the $L_{cl}$ loss function.}

\subsubsection{Determining the Temperature Constant}
\begin{figure}
\begin{center}
\includegraphics[scale=0.294]{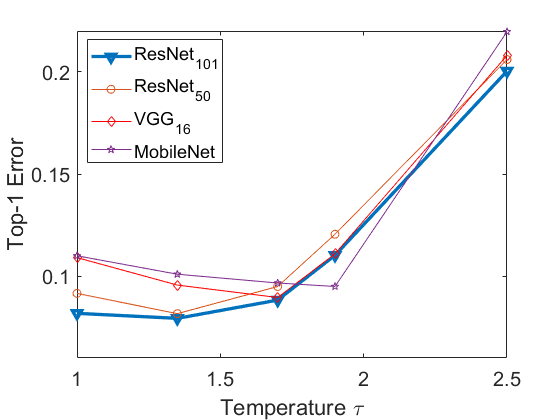}
\includegraphics[scale=0.294]{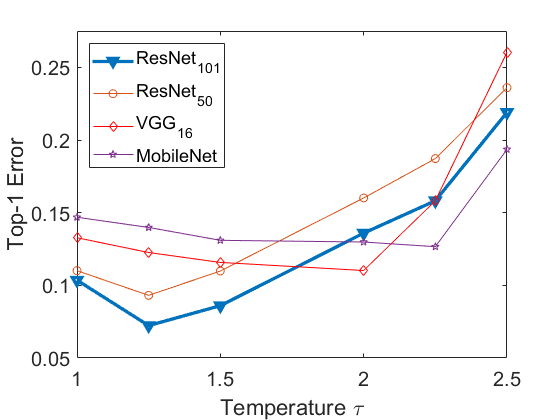}
\includegraphics[scale=0.294]{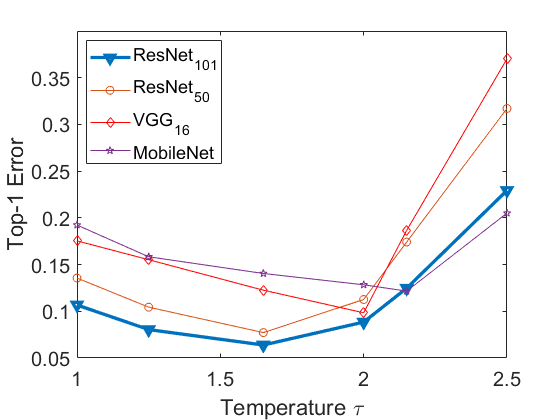}
\includegraphics[scale=0.294]{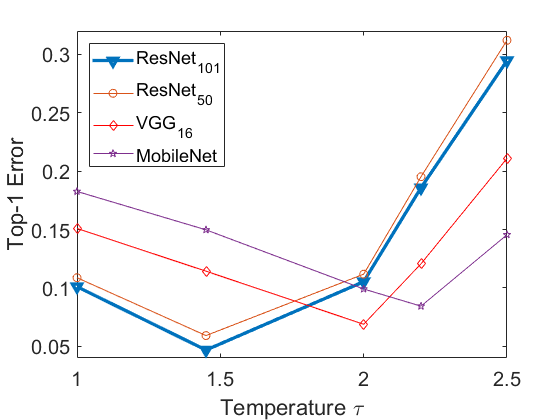}
\includegraphics[scale=0.294]{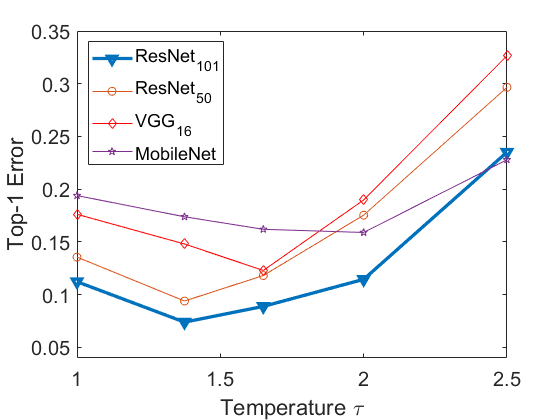}
\includegraphics[scale=0.294]{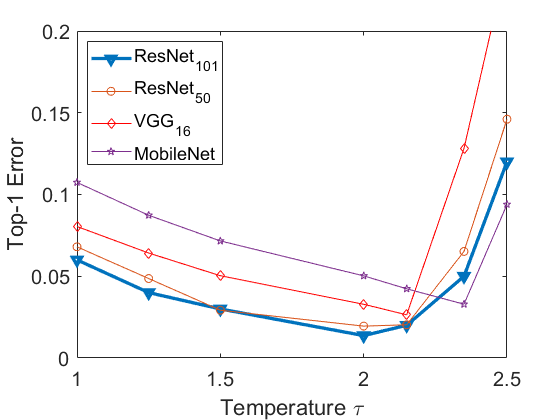}
\end{center}
   \caption{Effect of $\tau$ in terms of Top-1 error on: First row (left to right) Rabbani dataset, and the BIOMISA dataset. Second row (left to right): Duke-I+II dataset, and Duke-III dataset. Third row (left to right), Zhang dataset, and the proposed incremental cross-domain adaptation.}
\label{fig:short8}
\end{figure}

\noindent The temperature constant ($\tau$) is a hyperparameter that generates soft target probabilities for each class, thus, enabling the deep neural networks \RV{to learn these classes accurately during the knowledge distillation process.} $\tau$ is a dataset-dependent parameter, and its optimal value varies significantly across different domains. In this paper, we empirically determined the best value of $\tau$ for each dataset by analyzing the Top-1 classification error as shown in Figure \ref{fig:short8}. Here, the best value of $\tau$ differs for each dataset and each pre-trained model, but it typically ranges between $1 < \tau < 2.5$. For instance, on Rabbani dataset, the best classification performance for ResNet-101 \cite{resnet} was achieved with $\tau =$ 1.3. On the BIOMISA dataset, the optimal $\tau$ value is 1.25. Similarly, the optimal $\tau$ for the Zhang dataset and for the incremental cross-domain adaptation is 1.4 and 2, respectively (as shown in Figure \ref{fig:short8}). \RVVV{In addition to this, we also analyzed the performance of each network by varying $\tau$ on fixed intervals (across each dataset), and here we also found ResNet-101 to be the best classification model. More details on the effect of $\tau$ at fixed intervals are reported in the paper's supplementary material.}

\begin{table}[t]
\footnotesize
    \centering
    \caption{Comparison of different pre-trained models (incrementally trained using the proposed $L_{cl}$ function) to perform cross-domain retinopathy screening tasks. Bold indicates the best scores while the second-best scores are in blue color.}
    \begin{tabular}{c c c c c}
        \hline\small
         Metric &  MobileNet & ResNet-50  & ResNet-101  & VGG-16 \\\hline
         Acc &	0.9618&	\color{blue}{0.9678}& 	\textbf{0.9826}& 	0.9532\\ 
         TPR&	0.9490&	\color{blue}{0.9537}&	\textbf{0.9757}&	0.9299\\
         TNR&	0.9788&	\color{blue}{0.9864}&	\textbf{0.9918}&	0.9839\\
         PPV&	0.9834&	\color{blue}{0.9893}&	\textbf{0.9937}&	0.9870\\
         F1&	0.9659&	\color{blue}{0.9712}&	\textbf{0.9846}&	0.9576\\     
        \hline
    \end{tabular}
    \label{tab:models}
\end{table}

\subsubsection{Choice of the Classification Model}
\noindent This ablation study is related to determining the optimal classification network to perform the cross-domain retinopathy screening tasks. For this purpose, we utilize models such as MobileNet \cite{mobilenet}, ResNet-50 \cite{resnet}, ResNet-101 \cite{resnet}, and VGG-16 \cite{vgg}. The comparison is reported in Table \ref{tab:models}, where we can see that the best performance is achieved for ResNet-101. Moreover, the second-best performance is achieved by ResNet-50 across all the metrics. This is due to the fact that ResNets employ a residual feature fusion mechanism within the encoder part that allows the network to retain finer feature representations of the candidate scan during latent vector generation \cite{resnet}. However, in terms of computational speeds, ResNets might not be the optimal choice as they are not built to be lightweight like MobileNets. But since our prime objective in this study is to achieve better performance, we chose ResNet-101 in the rest of the experimentations.

\subsection{Comparison with the Conventional Methods}
\noindent In the first series of experiments, we measured the classification performance of the incremental ResNet-101 (trained using the proposed $L_{cl}$ loss function) on Rabbani, BIOMISA, Duke-III, and Zhang datasets. We used only these datasets here because 1) they are specifically designed for the retinal classification tasks, and 2) these datasets contain detailed scan-level clinicians' markings, which researchers have extensively used as ground truths for evaluating their methods \cite{LACNN, aoctnet, stgs}. Furthermore, to make the comparison fair, we only diagnosed the originally marked pathologies (within each dataset) for this experiment. For example, the proposed framework has been trained to extract normal, DME, CNV, and drusen pathologies from the OCT scans of the Zhang dataset. Similarly, the proposed framework extracts the normal, DME, and AMD pathologies from Rabbani and Duke-III datasets. Moreover, on the BIOMISA dataset, the proposed framework has been incrementally trained to extract the normal, ME, dry AMD, wet AMD, and CSR pathologies from the OCT scans. Here, we also excluded the evaluation on fundus scans because, to the best of our knowledge, there is no competitive framework that uses fundus scans from Rabbani and BIOMISA datasets to screen retinopathy.

\noindent The comparison of the proposed framework with state-of-the-art methods is reported in Table \ref{tab:comp}. Here, we can see that despite its incremental nature, the proposed framework outperforms its competitors by \RVV{1.86\%} on Zhang dataset in terms of accuracy, and \RVV{1.65\%} on Duke-III dataset, and \RVV{0.10\%} on BIOMISA datasets in terms of F1 score. In addition to this, the proposed framework is outperforming LACNN \cite{LACNN} on the Zhang dataset by \RVV{14.23\%} in terms of the F1 score, which is quite a noticeable improvement. \RVVV{Although, the STGS approach (proposed in \cite{stgs}) is outperforming the proposed framework by 0.308\% and 2.98\% in terms of Acc and TPR, respectively. However, we also wanted to highlight here that STGS is a conventional machine learning approach that has been trained on 3,840 scans. But the proposed framework, when trained on 1,340 scans, is able to produce competitive performance with STGS on the BIOMISA dataset (e.g., see the performance comparison of the proposed framework and STGS in terms of PPV and F1 scores within Table \ref{tab:comp}). Furthermore, since the STGS is a conventional machine learning framework, it lacks generalizability to process OCT scans acquired with different scanners simultaneously \cite{stgs}. }

\begin{table}[htb]
    \centering
    \caption{Performance comparison of the proposed framework (backboned through ResNet-101) with state-of-the-art retinopathy screening frameworks. To maintain fairness with the competitors, the proposed framework is incrementally trained only to identify the originally marked pathologies within each dataset. Moreover, bold indicates the best performance, whereas the second-best score is underlined. '-' indicates that the following metric is not computed by the other frameworks. The rest of the abbreviations are \RVVV{DT}: Dataset, Met: Metric, PF: Proposed Framework, \RVVV{CBRF: CNN-Based Referral Framework \cite{zhang}, HSVM: SVM driven by HOG Descriptor \cite{duke3}, AOCT: AOCT-Net \cite{aoctnet}, LACNN: Lesion-Aware CNN, STGS: Structure Tensor Graph Search \cite{stgs},} ZD: Zhang Dataset, RA: Rabbani, BO: BIOMISA, D3: Duke-III dataset.}   
    \begin{tabular}{cccccccccc}
    \hline
        \RVVV{DT} & Met & PF & \RVVV{CBRF} & \RVVV{HSVM} & \RVVV{AOCT} & \RVVV{LACNN} & \RVVV{STGS} \\\hline
        ZD & Acc &	\textbf{0.984} & \color{blue}{0.966} & - & 0.964 & 0.901 & -\\
        & TPR &	\textbf{0.981} & \color{blue}{0.978} & - &  0.962 & 0.868 & -  \\
        & TNR &	\textbf{0.984} & \color{blue}{0.974} & - &  0.968 & - & -  \\
        & PPV &	\textbf{0.994} & \color{blue}{0.994} & - &  0.989 & 0.862 & -  \\
        & F1 &	\textbf{0.987} & \color{blue}{0.986} & - & 0.975 & 0.864 & - \\\hline
        
        RA & Acc &	\textbf{0.972} & - & - & -  & - & - \\
        & TPR &	\color{blue}{0.968} & - & - & -  & \textbf{0.993} & - \\
        & TNR &	\textbf{0.980} & - & - & - & - & - \\
        & PPV &	\color{blue}{0.988} & - & - & -  & \textbf{0.993} & - \\
        & F1 &	\color{blue}{0.977} & - & - & -  & \textbf{0.993} & - \\\hline
        
        D3 & Acc &	\textbf{0.980} & - & \color{blue}{0.955} & - & - & -  \\
        & TPR &	\textbf{1.000} & - & \textbf{1.000} & - & - & -  \\
        & TNR &	\textbf{0.933} & - & \color{blue}{0.866} & - & - & -  \\
        & PPV &	\textbf{0.967} & - & \color{blue}{0.937} & - & - & -  \\
        & F1 &	\textbf{0.983} & - & \color{blue}{0.967} & - & - & - \\\hline 
        
        BO & Acc &	\color{blue}{0.974} & - & - & - & - & \textbf{0.977} \\
        & TPR &	\color{blue}{0.971} & - & - & - & - & \textbf{1.000} \\
        & TNR &	\textbf{0.996} & - & - & - & - & \color{blue}{0.933} \\
        & PPV &	\textbf{0.999} & - & - & - & - & \color{blue}{0.967} \\
        & F1 &	\textbf{0.984} & - & - & - & - & \color{blue}{0.983} \\
    \hline
    \end{tabular}
    \label{tab:comp}
\end{table}

\subsection{Incremental Cross-Domain Retinopathy Screening} \label{sec:combined}
\begin{figure}[t]
\begin{center}
    \includegraphics[width=1\linewidth]{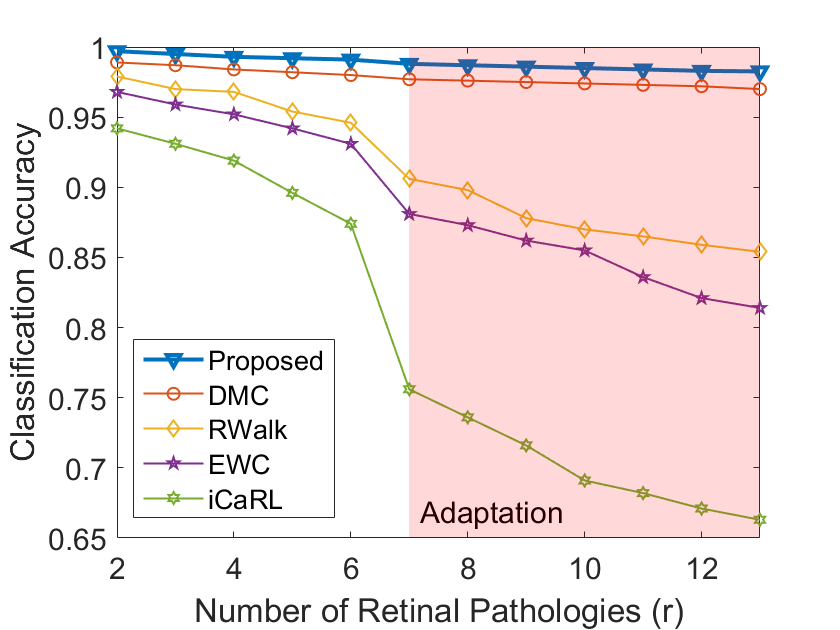}
\end{center}
   \caption{Performance comparison of the incremental ResNet-101 \cite{resnet} (driven through proposed $L_{cl}$) with iCaRL \cite{_8}, RWalk \cite{rwalk}, EWC \cite{ewc}, and DMC \cite{dmc} for cross-domain retinopathy screening.}
\label{fig:short999}
\end{figure}

\noindent In this series of experiments, we measured the capacity of the $L_{cl}$ to constrain the ResNet-101 in learning the retinopathy classification tasks simultaneously from the OCT and fundus imagery. The training details for the proposed incremental cross-domain adaptation are already discussed in Section \ref{sec:trainingDetails}. Moreover, its evaluation, in terms of classification accuracy, is reported in Figure \ref{fig:short999} and Table \ref{tab:domain}. In Figure \ref{fig:short999}, we can observe that as we continue learning retinal classification tasks, the $L_{cl}$ enables the ResNet-101  model to produce high resistance to catastrophic forgetting as compared to its competitors, such as DMC \cite{dmc}, RWalk \cite{rwalk}, EWC \cite{ewc} and iCaRL \cite{_8}. For example, when $r=13$, the proposed framework achieves \RVV{1.03\%} performance improvement compared to the second-best DMC-based ResNet-101 in terms of accuracy. These improvements stem from the fact that $L_{cl}$ not only constrains the classification models to retain their prior knowledge through distillation but also ensures that the classification models learn the structural and semantic relationships between incrementally learned representations.  Table \ref{tab:domain} showcases the performance comparison of the $L_{cl}$ driven ResNet-101 over its competitors for different domain adaptation configurations. Here, we can see that when the incremental ResNet-101 (trained using $L_{cl}$) is incrementally adapted from OCT to fundus imagery or from fundus to OCT imagery, it produces \RVV{1.03\%} and \RVV{3.22\%} better classification performance than the second-best DMC loss function, respectively towards recognizing different multi-modal retinal abnormalities. \NW{In Table \ref{tab:domain}, we can also notice that the proposed framework achieves \RVV{2.08\%} improvement when adapting from OCT to fundus domain as compared to the other configuration. This is because the training images for the OCT domain are relatively larger in nature than the fundus domain (see Table \ref{tab:data}, \ref{tab:data2}), and training the underlying network on OCT images first gives it more experience of screening the retinal pathologies, which leads to better classification performance at the inference stage.} 

\begin{figure}[t]
\begin{center}
    \includegraphics[width=1\linewidth]{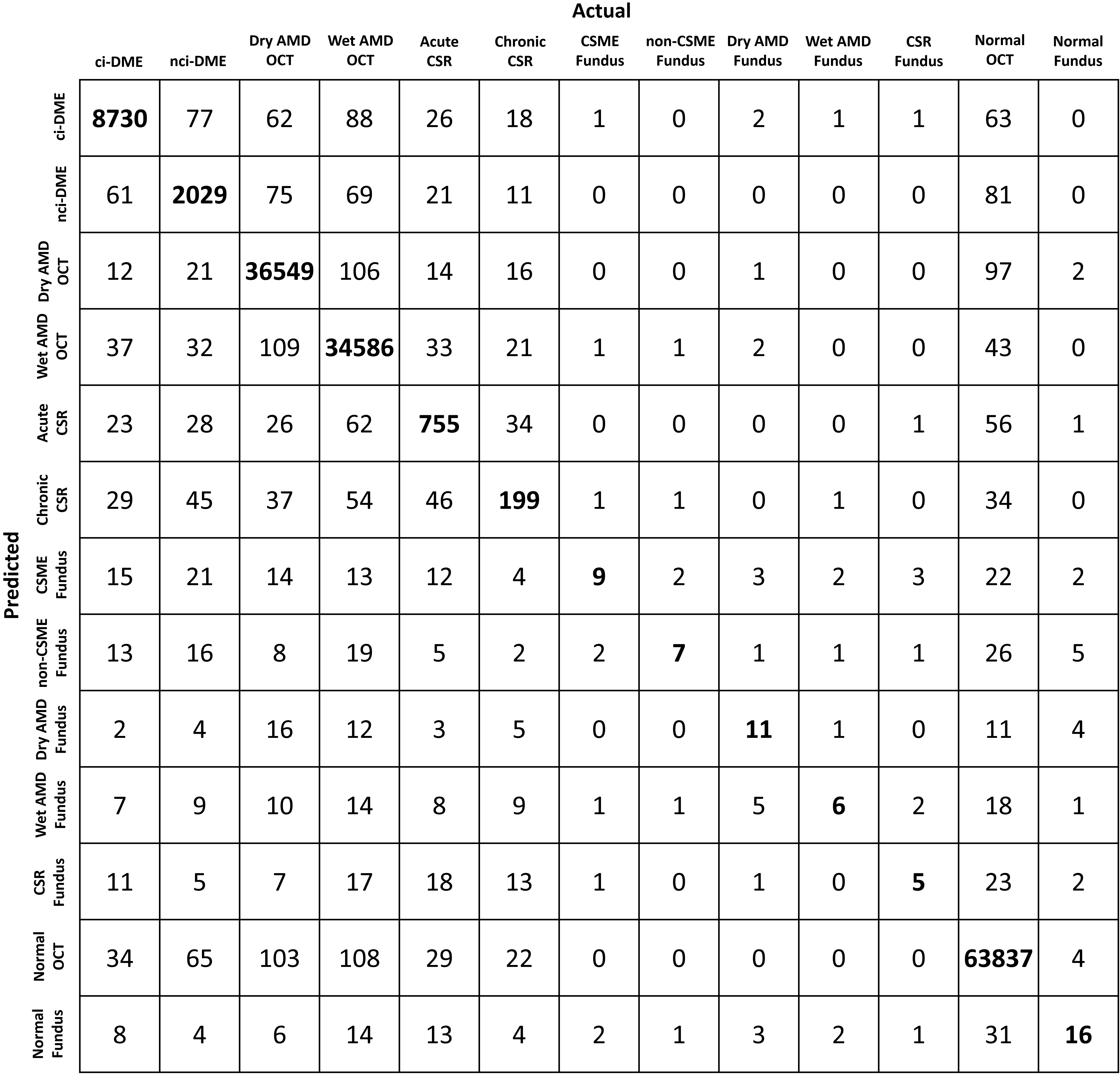}
\end{center}
   \caption{Confusion matrix depicting the performance of the incremental ResNet-101 (trained via proposed $L_{cl}$ loss function) to perform cross-domain retinopathy screening tasks. 
   }
\label{fig:short99}
\end{figure}

\noindent Apart from this, the classification performance of the proposed scheme for screening retinopathy from multiple modalities is also shown in the confusion matrix (Figure \ref{fig:short99}). We can notice here that regardless of the diversity within the scanner specifications, the incremental  $L_{cl}$-driven ResNet-101  achieved \RV{an} accuracy of 0.9826 and an F1 score of 0.9846. 

\begin{figure}[t]
\begin{center}
    \includegraphics[width=1\linewidth]{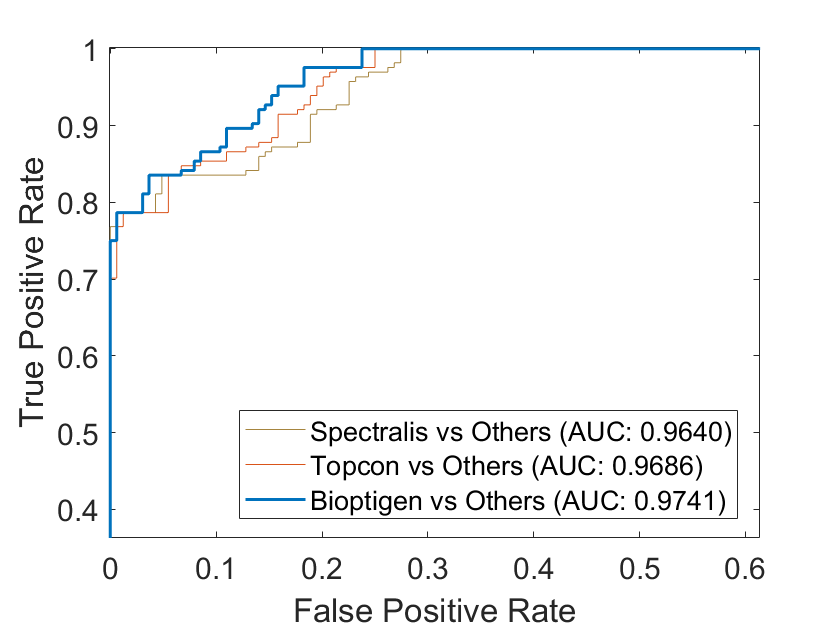}
\end{center}
   \caption{ROC curves showcasing the performance of the incremental ResNet-101 \cite{resnet} model (trained using $L_{cl}$ loss function) for screening the scanner-specific retinal pathologies.}
\label{fig:roc}
\end{figure}

\noindent \RVV{We further analyzed the capacity of the proposed framework (in terms of ROC curves) for discriminating scanner-specific retinal pathologies. For this experiment, we first categorized the fundus, and OCT scans from all the datasets w.r.t the machine manufacturers and then classified the pathologies from these scans using the proposed system. For example, for the curve representing 'Spectralis vs. Others', we have grouped the OCT and fundus scans from the Zhang, Rabbani, Duke-II, and Duke-III datasets (which are acquired through Spectralis, Heidelberg Inc. scanner). Similarly, we have also grouped the rest of the scans in the other datasets as 'Others'. Then, we checked whether or not the proposed framework correctly identified the retinal pathologies (within the Spectralis scans). The correct predictions are marked as 1's, whereas the misclassifications are marked as 0's.
Moreover, the correct predictions from other scanners were marked as 0's, and the misclassifications are marked as 1's. After labeling the predictions, we used their confidence scores and the ground truth labels to generate the 'Spectralis vs. Others' curve. The same process is repeated for generating the curves of other scanners as well. 

\noindent The ROC curves are demonstrated in Figure \ref{fig:roc}. Here, we can see that the proposed framework's classification performance, in terms of AUC scores, remains similar across different types of scanners. This evidences that the proposed framework remains unbiased towards screening retinal pathologies from any specific type of scanner during incremental cross-domain adaptation.}

\noindent In another series of experiments, we compared the cross-domain adaptation capacity of the proposed framework with the state-of-the-art schemes such as Fourier Domain Adaptation (FDA) \cite{fda}, Incremental Evolving Domain Adaptation (EDA) \cite{Bitarafan2020TKDE}, and Incremental Domain Adaptation for Neural Networks (iDANN) \cite{incDA2020arXiv}. 
The comparison is reported in Table \ref{tab:domainB}, where we can see that the proposed $L_{cl}$ driven ResNet-101 leads the second-best iDANN scheme by \RVV{2.08\%} and \RVV{4.34\%}, respectively, across both domain adaptation configurations. \TT{We also highlight that FDA is originally designed as a domain adaptation scheme for semantic segmentation tasks \cite{fda}. However, to fairly compare it with the proposed framework, we utilized it for classification purposes, i.e., we first stylized the training scans using FDA and then used them to incrementally train the ResNet-101 via the proposed $L_{cl}$ loss function. After training the model, we evaluated it for retinopathy screening using the FDA stylized OCT and fundus test scans. Here, the stylization, through FDA, is meant to overcome the scanner differences, which should aid the underlying network towards accurately learning the cross-modal classification tasks. But from Table \ref{tab:domainB}, we can see that the classification performance of the incrementally trained ResNet-101 on the FDA stylized scans is not the best. The low performance of \RV{the} FDA scheme stems from the fact that it uses sharp cut-off transitions of the parametric-rectangular window for stylization \cite{fda}, which also introduces a large number of noisy transitions within the transformed scans. Also, it should be noted that the FDA scheme is primarily designed for the inter-related semantic segmentation tasks only \cite{fda}.

\noindent Apart from this, we can notice that the classification performance of the proposed scheme and the iDANN is comparable for both configurations (see Table \ref{tab:domainB}). iDANN incrementally learns the pool of experiences from the scarce (yet inter-related) domains via gradient reversal strategy \cite{incDA2020arXiv}, where the underlying classification model is trained to recognize the domain-invariant features from the subset of target domain training samples (chosen in an unsupervised manner). This eventually achieves higher generalizability for performing cross-domain tasks simultaneously \cite{incDA2020arXiv}. However, while adapting from Target Domain-II to Target-Domain-I (i.e., from \RV{the} fundus to OCT imagery), the performance of iDANN deteriorates at a higher rate than the $L_{cl}$ driven incremental cross-domain adaptation scheme (as evident from Table \ref{tab:domainB}). This is because $L_{cl}$ analyzes the structural and semantic similarities, via Bayesian inference, between incrementally learned knowledge representations, while iDANN only focuses on the gradients minimization (during the back-propagation) based upon the spatial feature representations \cite{incDA2020arXiv}.  
}

\begin{table}[htb]
    \centering
    \caption{\NW{Performance comparison of $L_{cl}$ loss function with state-of-the-art incremental learning schemes (in terms of accuracy) for performing the cross-domain retinopathy screening tasks. The source represents the ImageNet dataset \cite{imagenet}. Target-I represents the OCT imagery taken from Rabbani, BIOMISA, Zhang, Duke-I, Duke-II, and Duke-III \RV{datasets}. Target-II represents the fundus imagery taken from Rabbani and BIOMISA dataset. Moreover, the performance comparison is measured using \RV{the} ResNet-101 model. The best performance is in bold, while the second-best performance is in blue color. The rest of the abbreviations are iC: iCaRL \cite{_8}, EC: EWC \cite{ewc}, RW: RWalk \cite{rwalk}, and DC: DMC \cite{dmc}.}}
    \begin{tabular}{ccccccc}
    \hline
        Configuration & $L_{cl}$ & iC & EC & RW & DC \\\hline
        Source: Target-I $\rightarrow$ Target-II & \textbf{0.98} & 0.66 & 0.80 & 0.85 & \color{blue}{0.97} \\
        Source: Target-II $\rightarrow$ Target-I & \textbf{0.96} & 0.68 & 0.81 & 0.83 & \color{blue}{0.93} \\
        
    \hline
    \end{tabular}
    \label{tab:domain}
\end{table}

\begin{table}[htb]
    \centering
    \caption{\NW{Performance comparison of the proposed framework with state-of-the-art incremental domain adaptation approaches in terms of classification accuracy. For a fair comparison, we used ResNet-101 in all the schemes for classification purposes. Bold indicates the best performance while the second-best performance is in blue color. Moreover, the abbreviations are SD: Source Domain (ImageNet \cite{imagenet}), T-I: Target Domain-I, and T-II: Target Domain-II.}}
    \begin{tabular}{cccccc}
    \hline
        Configuration & Proposed & FDA \cite{fda} & EDA \cite{Bitarafan2020TKDE} & iDANN \cite{incDA2020arXiv}\\\hline
        SD: T-I $\rightarrow$ T-II & \textbf{0.98} & 0.86 & 0.89 & \color{blue}{0.96}\\
        SD: T-II $\rightarrow$ T-I & \textbf{0.96} & 0.81 & 0.84 & \color{blue}{0.92}\\
        
    \hline
    \end{tabular}
    \label{tab:domainB}
\end{table}

\section{Discussion} \label{sec:discussion}
\begin{figure}[t]
\begin{center}
\includegraphics[width=1\linewidth]{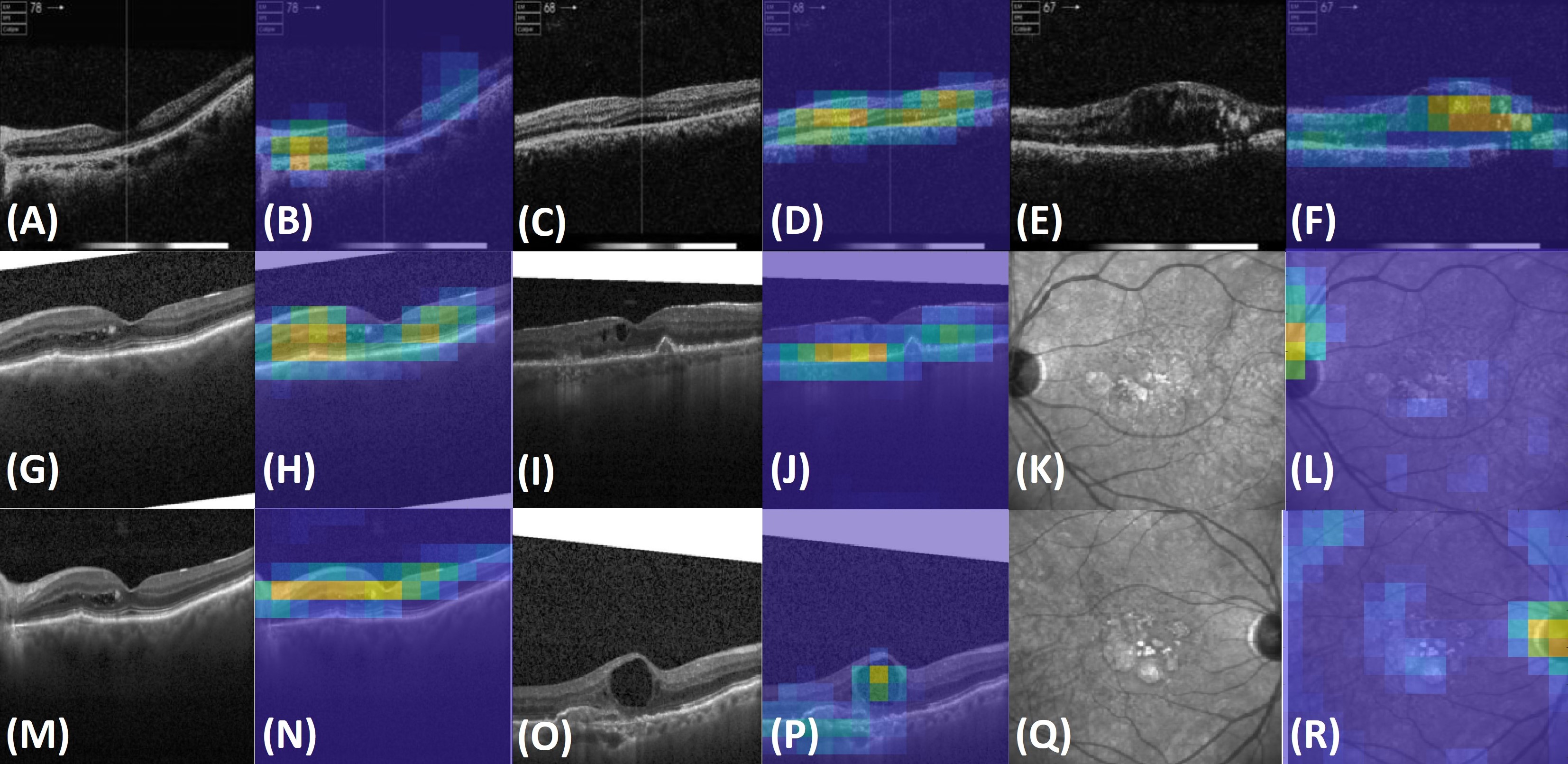}
\end{center}
   \caption{Attention maps obtained from the incremental ResNet-101 (trained via proposed $L_{cl}$ loss function) when performing the cross-domain retinopathy screening tasks. }
\label{fig:short5}
\end{figure}

\noindent 
After rigorously evaluating the proposed framework on the six public datasets against different experimental settings, the proposed system showed an overall superiority over state-of-the-art retinal diagnostic frameworks. In addition to this, the results, reported in Tables \ref{tab:domain}, \ref{tab:domainB} and Figure \ref{fig:short999} evidences a neat improvement of the proposed framework over state-of-the-art incremental learning, domain adaptation, and incremental domain adaptation schemes for performing the cross-domain retinopathy screening tasks. \RV{Furthermore, we have thoroughly tested the applicability of the proposed framework for real-time retinopathy screening through blind testing experiments in clinical settings. Due to space constraints, we reported these experiments within the supplementary material of the paper.} \NW{Apart from this, our framework's core element is the $L_{cl}$ loss function, in which the newly proposed mutual distillation loss $L_{md}$ plays a capital role. The importance of $L_{md}$ objective in $L_{cl}$ can also be seen in Table \ref{tab:my_label11} (for all the datasets) where the $L_{md}$ objective function, driven through Bayesian inference, significantly aids in learning the new classification categories while simultaneously performing well on the already known tasks.}
\begin{table}[t]
\footnotesize
    \centering
    \caption{Performance comparison of $L_{cl}$ (with and without $L_{md}$) on different datasets. Bold indicates the best performance. Moreover, the abbreviations are RA: Rabbani dataset, BO: BIOMISA dataset, DI: Duke I+II dataset, D3: Duke-III dataset, ZD: Zhang dataset, ICDA: Incremental Cross-Domain Adaptation.} 
    \begin{tabular}{c c c c c c c}
        \hline\small
        Loss Functions&	RA &	BO &	DI  &	D3 &	ZD & ICDA\\
        \hline
 $L_{cl}$ with $L_{md}$ &	\textbf{0.972}&	\textbf{0.974}&	\textbf{0.963}&	\textbf{0.980}&	\textbf{0.984}&	\textbf{0.982}\\
 $L_{cl}$ w/o $L_{md}$ &0.856&	0.859&	0.847&	0.894&	0.886&	0.764\\
        \hline
    \end{tabular}
    \label{tab:my_label11}
\end{table}
\noindent \R{We also want to highlight here that although $L_{cl}$ is robust in avoiding the catastrophic forgetting during the incremental training (as evidenced from Section \ref{sec:results}), we did notice its vulnerability towards screening retinopathy when the classes are imbalanced. For example, in Figure \ref{fig:short99}, we can see that the proposed framework (trained with $L_{cl}$) correctly screened 36,549 out of 37,022 dry AMD pathologies, resulting in the classification score of 0.9872. However, in fundus scans, 11 dry AMD cases were classified correctly out of 29. This significant difference in the classification performance is due to the imbalanced number of the fundus and OCT scans on which network was trained, i.e., the quantity of fundus scans is very small compared to OCT scans (see Table \ref{tab:data} and \ref{tab:data2} for more details).
Therefore, the classification model within the proposed framework is more biased towards accurately screening the retinal pathologies from the OCT scans. \RVV{Also, notice the differences in the attention maps between OCT and fundus scans (within Figure \ref{fig:short5}). 
These attention maps show the abnormal regions which the network focuses on while predicting the underlying pathology, and by observing the pairs in Figure \ref{fig:short5} (A-B), (C-D), (E-F), (G-H), (I-J), (M-N), (O-P), we can see that the proposed system correctly recognizes the retinal lesions such as choroidal neovascularization, sub-retinal fluid, intra-retinal fluid, hard exudates, drusen from the dry and wet AMD, acute and chronic CSR, ci-DME and nci-DME pathologies. However, if we look at the pairs (K-L) and (Q-R) in Figure  \ref{fig:short5}, we can note that the attention of the proposed system is more focused on the optic-disc region rather than the retinal lesions. This biasness (in the attention maps) is due to the highly imbalanced ratio of the OCT and fundus scans in the datasets (see Table \ref{tab:data2}). \R{But it should be noted at the same time that, clinically, the bias of the proposed framework towards OCT imagery is also very significant because the OCT imagery can objectively show the symptomatic appearance of the retinal abnormalities in early stages as compared to the fundus scans \cite{review}, which leads towards timely diagnosis for the prevention of the candidate's vision loss. Nevertheless, we can overcome these imbalanced cases by introducing some class imbalance remedies using max-margin constraints \cite{maxmargin}, focal loss function \cite{retinanet}, and Gaussian affinity optimizations \cite{affinity}.}}
\RV{\noindent Apart from this, currently, we do not use any denoising method to enhance the quality of the scans. The reasons for not using this preprocessing step are: 1) the proposed system employs CNN backbones to generate latent vectors which implicitly remove the high-frequency (noisy) content during scan decomposition, and the denoising step is generally required for the segmentation frameworks as they also need to reconstruct the segmented pixel-level maps back where noisy pixels in the input scan may lead to large pixel-level false positives \cite{ragfw}. 2) The classification performance of the proposed framework in terms of accuracy and F1 scores is already very reasonable, i.e., 0.9826 and 0.9846, respectively. Nevertheless, we also agree that adding a denoising step (involving wiener filtering \cite{syed2016CMPB} or structure tensors \cite{tbme}) would further increase the classification performance. But since the performance of the proposed framework is already decent, we deemed this as an optional step that can be opted in the future. 
}
}

\section{Conclusion} \label{sec:conclusion}
\noindent This paper presents a novel incremental cross-domain adaptation approach that enables the classification networks to retain their prior knowledge while learning new cross-modality retinopathy screening tasks incrementally via few-shot training. The proposed scheme offers a highly scalable option for the clinical settings, aiding the ophthalmologists in screening the vast majority of retinal abnormalities (even the rarely occurring ones) with few training examples. The extent of the proposed scheme has been tested with different pre-trained models on six publicly available datasets across two domains, where it outperforms state-of-the-art incremental learning and domain adaptation schemes for identifying thirteen types of retinal pathologies simultaneously. 
\noindent The backbone of the proposed scheme is the $L_{cl}$ loss function which \RVV{constrains} the classification networks, via Bayesian inference, to gain higher generalizability for performing the diversified cross-domain tasks. Furthermore, due to the promising performance of the proposed scheme for screening retinal diseases across multi-modal imagery,  we envisage testing its applicability on other medical applications in the future. \R{Moreover, in the future, we also anticipate overcoming the vulnerability of $L_{cl}$ loss function against imbalanced classes by introducing the max-margin constraints and modulating factors that will boost the capacity of the classification models (trained on $L_{cl}$) to predict the imbalanced classes accurately.} 

\section*{Acknowledgement}
\noindent This work is supported by a research fund from Khalifa University: Ref: CIRA-2019-047, and the Abu Dhabi Department of Education and Knowledge (ADEK), Ref: AARE19-156. \RV{We would also like to thank all the ophthalmologists from the Armed Forces Institute of Ophthalmology (AFIO), Rawalpindi, Pakistan, for thoroughly annotating the retinal scans in all six datasets, and aiding us in the blind testing process.}

\footnotesize
\bibliographystyle{ieeetr}


\end{document}